\documentclass{article}

\usepackage{natbib}
\usepackage{graphicx}

\usepackage{bm}% bold math

\makeatletter
\long\def\@makecaption#1#2{%
   \vskip 10\p@
   \setbox\@tempboxa\hbox{{\bf #1: }#2}%
   \ifdim \wd\@tempboxa >\hsize
       {\bf #1:}#2\par
     \else
       \hbox to\hsize{\hfil\box\@tempboxa\hfil}%
   \fi}
\makeatother

\newcommand{\latexOrPdflatex}[2]{\ifx\undefined\pdfoutput%
#1%
\else%
#2%
\fi}

\latexOrPdflatex{
\newcommand{\href}[2]{#2}
}{
\usepackage[colorlinks,pagebackref,citecolor=blue,bookmarksopen,bookmarksnumbered,pdftitle={},pdfauthor={Wiskott},a4paper]{hyperref}
}

\renewcommand{\vec}[1]{\bm{#1}}	% vector boldface
\newcommand{\mat}[1]{\bm{#1}}	% matrix boldface

\newcommand{\is}{x}			% input signal
\newcommand{\tfc}{h}			% transfer function component
\newcommand{\es}{z}			% expanded signal
\newcommand{\pr}{a}			% parameters of the transfer function
\newcommand{\tf}{g}			% transfer function
\newcommand{\os}{y}			% output signal

\newcommand{\iu}{i}			% sample index
\newcommand{\uu}{w}			% time series
\newcommand{\p}{\gamma}			% true driving force
\newcommand{\pest}{\gamma^{est}}	% estimated driving force

\newcommand{\emb}{x}			% embedding vector
\title{%
\vspace{-0.5cm}{\small \tt \centerline{e-print published at
\href{http://www.arxiv.org/abs/cond-mat/0312317}{http://www.arxiv.org/abs/cond-mat/0312317} December 2003}}
\vspace{1.cm} 
Estimating Driving Forces of Nonstationary Time Series \\ with Slow Feature
Analysis}
\author{Laurenz Wiskott \\ \\
 Institute for Theoretical Biology, Humboldt-University Berlin \\
 Invalidenstra{\ss}e~43, D-10115 Berlin, Germany \\
 \texttt{\href{http://itb.biologie.hu-berlin.de/~wiskott}{http://itb.biologie.hu-berlin.de/{\footnotesize$\sim$}wiskott}} \\
 \texttt{l.wiskott@biologie.hu-berlin.de}
}

\date{}

\begin{document} 

\maketitle

\begin{abstract}
 Slow feature analysis (SFA) is a new technique for extracting slowly
 varying features from a quickly varying signal.  It is shown here that SFA
 can be applied to nonstationary time series to estimate a single
 underlying driving force with high accuracy up to a constant offset and a
 factor.  Examples with a tent map and a logistic map illustrate the
 performance.
\end{abstract}

%PRL \keywords{driving force, nonlinear time series analysis, nonstationary time series, slow feature analysis}

\section{Introduction} \label{sec:introduction}

 Nonlinear time series analysis generally assumes stationarity
 \cite[see][for an overview]{Casd1997,Schr1999}.  However, many time series
 are actually nonstationary for various reasons, such as temperature drift
 in the experimental setup, decreasing reservoirs in (bio)chemical reactors
 or ecological systems, global warming in climate data, varying heart rate
 in cardiology, or vibrato/tremor in vocal fold vibrations.  Such
 nonstationarities can be modeled by underlying parameters, referred to as
 driving forces, that change the dynamics of the system smoothly on a slow
 time scale or abruptly but rarely, e.g.\ if the dynamics switches between
 different discrete states.

 If a test reveals that a time series is
 nonstationary~\cite[]{WittKurtPiko1998}, one can still apply methods of
 stationary time series analysis if one determines sections where the
 driving force has similar values and analyzes these sections as one
 stationary time series.  This can be done by first slicing the time series
 into windows of equal size and then grouping the windows based on
 similarity measures of the dynamics~\cite[]{ManuSavi1996}.
 This division of the time series can be avoided by the technique of
 overembedding.  If the embedding dimension is sufficiently high, then
 similar embedding vectors automatically belong to similar values of the
 driving forces and all available data can be used for analysis, such as
 forecasting or nonlinear noise reduction~\cite[]{HeggKantMata+2000}.

 However, in some cases, e.g.\ in the analysis of EEG data, one is
 particularly interested in revealing the driving forces themselves, which
 is in principle only possible up to an invertible transformation. 
 One standard method for visualizing driving forces is the recurrence
 plot~\cite[]{EckmOlifRuel1987}, but it is often difficult to interpret.
 Methods have been developed to estimate driving forces based on
 the finding that the recurrence plot of a time series is similar to the
 recurrence plot of its underlying driving forces~\cite[]{Casd1997,Schr1999}.
 Another technique for the reconstruction of driving forces has been
 presented in~\cite[]{VerdGranNavo+2001}.
 Here I present an alternative approach based on slow feature analysis, a
 new technique developed in the field of theoretical neurobiology.

\section{Slow Feature Analysis}

 Slow feature analysis (SFA) has been originally developed in context of an
 abstract model of unsupervised learning of invariances in the visual
 system of vertebrates~\cite[]{Wis98a} and is described in detail
 in~\cite[]{WisSej2002}.

 The general objective of SFA is to extract slowly varying features from a
 quickly varying signal.  For a scalar output signal it can be formalized
 as follows.  Let $\vec{\is} = \vec{\is}(t)$ be an $N$-dimensional input
 signal where $t$ indicates time and $\vec{\is} = [\is_1,...,\is_N]^T$ is a
 vector.  Find the input-output function $\tf(\vec{\is})$ that generates a
 scalar output signal
\begin{equation}
 \os(t) := \tf(\vec{\is}(t))
\end{equation}
 with most slowly temporal variation possible as measured by the variance
 of the time derivative:
\begin{equation}
 \mbox{minimize} \quad \langle\dot{\os}^2\rangle \label{eq:slowness}
\end{equation}
 with $\langle \cdot \rangle$ indicating the temporal mean.  For
 convenience and to avoid the trivial constant solution the output signal
 has to meet the following constraints:
%
%PRL \begin{subequations} \label{eq:constr0-constr1}
\begin{eqnarray}
 \langle\os\rangle &=& 0 \quad \mbox{(zero mean)} \,, \label{eq:constr0} \\
 \langle\os^2\rangle &=& 1 \quad \mbox{(unit variance)} \label{eq:constr1} \,.
\end{eqnarray}
%PRL \end{subequations}

 This is an optimization problem of variational calculus and as such
 difficult to solve.  However, if we constrain the input-output function to
 be a linear combination of some fixed and possibly nonlinear basis
 functions, the problem becomes tractable and can be solved in the
 following way.
 Let $\vec{\tfc}' = \vec{\tfc}'(\vec{\is})$ be a vector of some fixed basis
 functions.  To be concrete assume $\vec{\tfc}'$ contains all monomials of
 degree~1 and~2.  Applying $\vec{\tfc}'$ to the input signal $\vec{\is}(t)$
 yields the nonlinearly expanded signal $\vec{\es}'(t)$:
%
%PRL \begin{subequations}
\begin{eqnarray}
 \vec{\tfc}'(\vec{\is}) &:=& [\is_1,\, \is_2,\, ...,\, \is_N,\,
% \nonumber \\ &&
 \is_1^2,\, \is_1 \is_2,\, ...,\, \is_N^2]^T, \\
 \vec{\es}'(t) &:=& \vec{\tfc}'(\vec{\is}(t)) \,.
\end{eqnarray}
%PRL \end{subequations}

 Assume $\tf(\vec{\is})$ is a linear combination of the basis functions
 plus a constant $c$, generating the output signal
 $\os(t)=\tf(\vec{\is}(t))$:
%
%
%PRL \begin{subequations}
\begin{eqnarray}
 \tf(\vec{\is}) & := & c + {\vec{\pr}'}^T \vec{\tfc}'(\vec{\is}) \,, \\
 \os(t) & = & c + {\vec{\pr}'}^T \vec{\es}'(t) \,.
\end{eqnarray}
%PRL \end{subequations}
%
 Since $\vec{\tfc}'$ contains all monomials of degree~1 and~2,
 $\tf(\vec{\is})$ can be any polynomial of degree~2, if the constant $c$
 and the coefficient vector ${\vec{\pr}'}$ are chosen correspondingly.

 For reasons that become clear below, it is convenient to sphere (or
 whiten) the expanded signal and transform the basis functions accordingly:
%
%PRL \begin{subequations}
\begin{eqnarray}
 \vec{\tfc}(\vec{\is}) &:=& \mat{S} \ (\vec{\tfc}'(\vec{\is}) - \langle
 \vec{\es}' \rangle) \,, \\
 \vec{\es}(t) &:=& \mat{S} \ (\vec{\es}'(t) - \langle \vec{\es}'
 \rangle) \,,
\end{eqnarray}
%PRL \end{subequations}
%
 with the sphering matrix $\mat{S}$ chosen such that the signal components
 have a unit covariance matrix, i.e.\ $\langle \vec{\es} \vec{\es}^T
 \rangle = \mat{I}$, which can be easily done with the help of principal
 component analysis (PCA).  The signal components have also zero mean,
 $\langle \vec{\es} \rangle = \vec{0}$, since the mean values have been
 subtracted.

 The input-output function and the output signal can now be written in
 these transformed basis functions and sphered expanded signal,
 respectively:
%
%PRL \begin{subequations} \label{eq:g2-y2}
\begin{eqnarray}
 \tf(\vec{\is}) &=& \vec{\pr}^T \vec{\tfc}(\vec{\is}) \label{eq:g2} \,, \\ 
 \os(t) &=& \vec{\pr}^T \vec{\es}(t) \label{eq:y2} \,.
\end{eqnarray}
%PRL \end{subequations}
%
 This is a form in which the constraints (\ref{eq:constr0}, \ref{eq:constr1}) are
 particularly easy to meet, since we find for any coefficient vector
 $\vec{\pr}$ with norm 1,
%
%PRL \begin{subequations}
\begin{eqnarray}
 \langle\os\rangle &=& \vec{\pr}^T
 \underbrace{\langle\vec{\es}\rangle}_{=\vec{0}} = 0 \,, \\
 \langle\os^2\rangle &=& \vec{\pr}^T \underbrace{\langle \vec{\es}
 \vec{\es}^T \rangle}_{= \mat{I}} \vec{\pr} = \vec{\pr}^T \vec{\pr} = 1 \,,
\end{eqnarray}
%PRL \end{subequations}
%
 which also motivates why constant $c$ has been dropped in
 (\ref{eq:g2}, \ref{eq:y2}).

 Thus the optimization problem reduces to finding the normalized
 coefficient vector $\vec{\pr}$ that minimizes
\begin{equation}
 \langle \dot{\os}^2 \rangle = \vec{\pr}^T \langle \dot{\vec{\es}}
 \dot{\vec{\es}}^T \rangle \vec{\pr} \,, \label{eq:yDot22}
\end{equation}
 which is obviously the normalized eigenvector of the time-derivative
 covariance matrix $\langle \dot{\vec{\es}} \dot{\vec{\es}}^T \rangle$ with
 the smallest eigenvalue, which again can be easily found by PCA.  The
 output signal $\os(t)$ is then given by (\ref{eq:y2}).  Notice that
 $\os(t)$ is usually uniquely determined up to the sign.

 In summary SFA consists of basically four steps: (i)~expand the input
 signal with some set of fixed possibly nonlinear functions; (ii)~sphere
 the expanded signal to obtain components with zero mean and unit
 covariance matrix; (iii)~compute the time derivative of the sphered
 expanded signal and determine the normalized eigenvector of its covariance
 matrix with the smallest eigenvalue; (iv)~project the sphered expanded
 signal onto this eigenvector to obtain the output signal.  The
 input-output function is given by (\ref{eq:g2}) but is not needed here,
 since we are only interested in the extracted output signal.

 In practice one has to work with time series $\vec{\is}_\iu$ instead of
 continuous signals $\vec{\is}(t)$, but the transfer of the algorithm
 described above to time series is straight forward.  The time derivative
 is simply computed as the difference between successive data points
 without wrap-around assuming a constant sampling spacing $\Delta t$.

 The description of SFA given above assumes that all monomials of degree~1
 and~2 are used for signal expansion.  However, any other set of basis
 functions could be used as well, e.g.\ monomials of higher degree or
 radial basis functions.
 It is also straight forward to extend SFA to more than one output signal
 component, in which case additional components should be uncorrelated to
 earlier ones.  Under this constraint additional components can be computed
 by using the normalized eigenvectors of (\ref{eq:yDot22}) with the next
 larger eigenvalues~\cite[]{WisSej2002}.
 SFA requires that all components vary to some extent.  Thus if there are
 constant components in the input signal or the expanded signal, these are
 simply discarded.

\section{Examples}

 In the following I will present two examples with time series $\uu_\iu$
 derived from a tent map and a logistic map to illustrate the properties of
 SFA.  I have also done simulations with the Lorenz System, but results
 were very unreliable and sensitive to parameter variations and noise.  The
 underlying driving force will always be denoted by $\p$ and may vary
 between $-1$ and $1$ either smoothly or rarely, but with a comparable time
 scale of variation (as defined by the variance of its time derivative
 (\ref{eq:slowness})).

 Taking the time series $\uu_\iu$ directly as an input signal would not give
 SFA enough information to estimate the driving force, because SFA
 considers only data (and its derivative) from one time point at a time.
 Thus it is necessary to generate an embedding-vector time series as an
 input.  Here embedding vectors at time point $\iu$ with delay $\tau$ and
 dimension $m$ are defined by
\begin{eqnarray}
 \vec{\emb}_\iu &:=& [\uu_{\iu - \tau (m-1)/2},\, \uu_{\iu - \tau
 ((m-1)/2-1)},\, ..., \uu_{\iu + \tau (m-1)/2}]^T \,,
\end{eqnarray}
 for scalar $\uu_\iu$ and odd $m$.  The definition can be easily extended to
 even $m$, which requires an extra shift of the indices by $\tau/2$ or its
 next lower integer to center the used data points at $\iu$.  Centering the
 embedding vectors results in an optimal temporal alignment between
 estimated and true driving force.

 The following simulations are based on 6000 data points each and were done
 with \textsc{Matlab} (Release 13).

\subsection{Tent Map}

\begin{figure*}[htbp!]

\centerline{\includegraphics[width=0.926\textwidth]{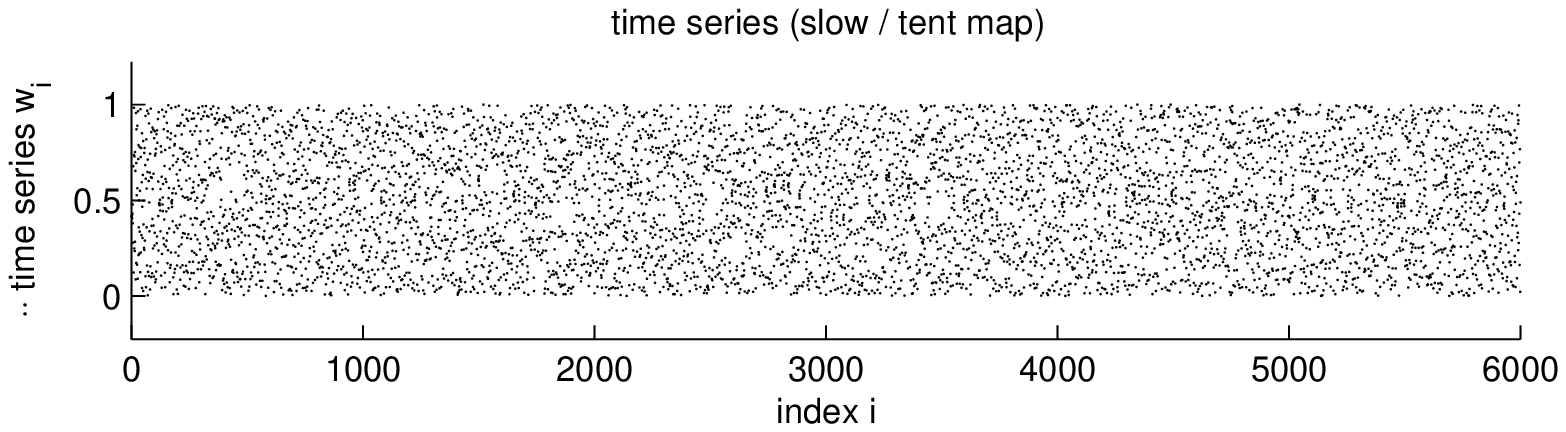}\hspace*{0.044\textwidth}}
\centerline{\includegraphics[width=0.97\textwidth]{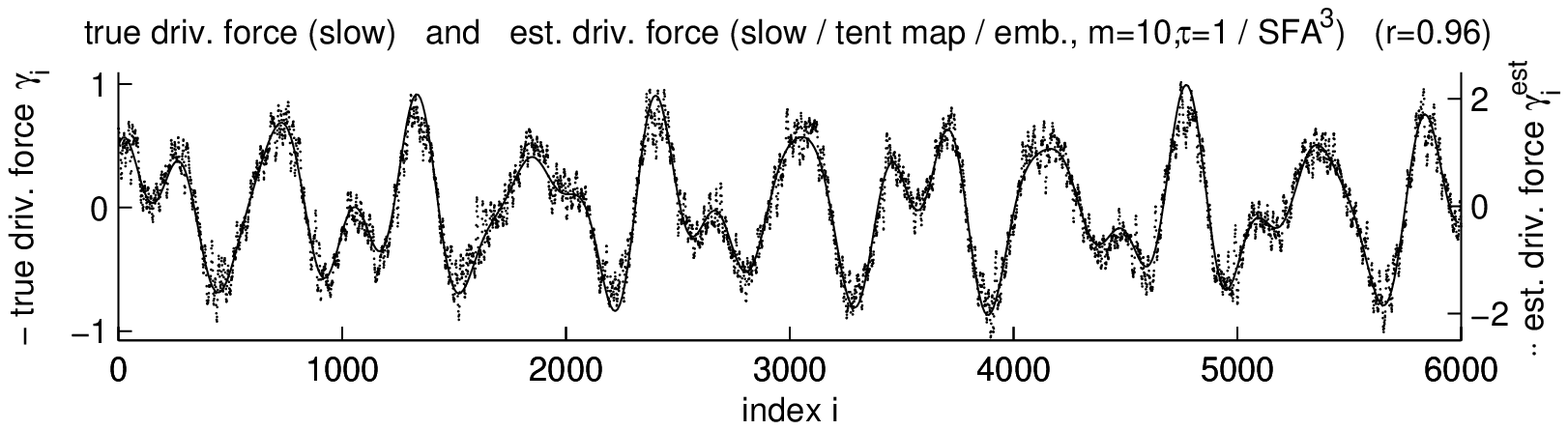}}

 \caption{\label{fig:SmoothTent} True slowly varying driving force $\p_\iu$
 (bottom, solid line), time series $\uu_\iu$ (top) derived from the driving
 force with the tent map, and estimated driving force $\pest_\iu$ (bottom,
 dots).  The correlation between true and estimated driving force is $r=0.96$.}

\end{figure*}

 As a first example consider a time series generated by an iterated tent
 map~\cite[]{Casd1997}
\begin{eqnarray}
 \hat{\p}_\iu &:=& (\p_\iu+1)/2 \\
 \uu_{\iu+1} &=& \left\{ \begin{array}{l@{\hspace{4ex}\mbox{if}\quad}r@{\,}c@{\,}l} 
 +2\uu_\iu+\hat{\p}_\iu & 0 \le &\uu_\iu& \le \frac{1}{2}(1-\hat{\p}_\iu) \\
 -2\uu_\iu-\hat{\p}_\iu+2 & \frac{1}{2}(1-\hat{\p}_\iu) < &\uu_\iu& \le
 \frac{1}{2}(2-\hat{\p}_\iu) \\
 +2\uu_\iu+\hat{\p}_\iu-2 & \frac{1}{2}(2-\hat{\p}_\iu) < &\uu_\iu& \le 1
 		     \end{array} \right. \,,
\end{eqnarray}
 which maps the interval $[0,1]$ onto itself with a functional form
 ``$\bigwedge$'' for $\p=-1$ (or $\hat{\p}=0$).  The parameter $\p$ shifts
 this wedge cyclically to the left until it becomes a ``$\bigvee$'' for
 $\p=\hat{\p}=+1$.  Figure~\ref{fig:SmoothTent} shows the true driving
 force, the time series, and the estimated driving force with $m=10$,
 $\tau=1$, and third order polynomials for SFA.  The correlation between
 true and estimated driving force is $r=0.96$.  Note that the scale and
 offset of the estimated driving force are arbitrarily fixed by the
 constraints and that the sign is random.  The axes were therefore chosen
 such that the curves in the bottom graphs of this and the following
 figures are optimally aligned with each other.

\subsection{Logistic Map}

\begin{figure*}[htbp!]

\centerline{\includegraphics[width=0.926\textwidth]{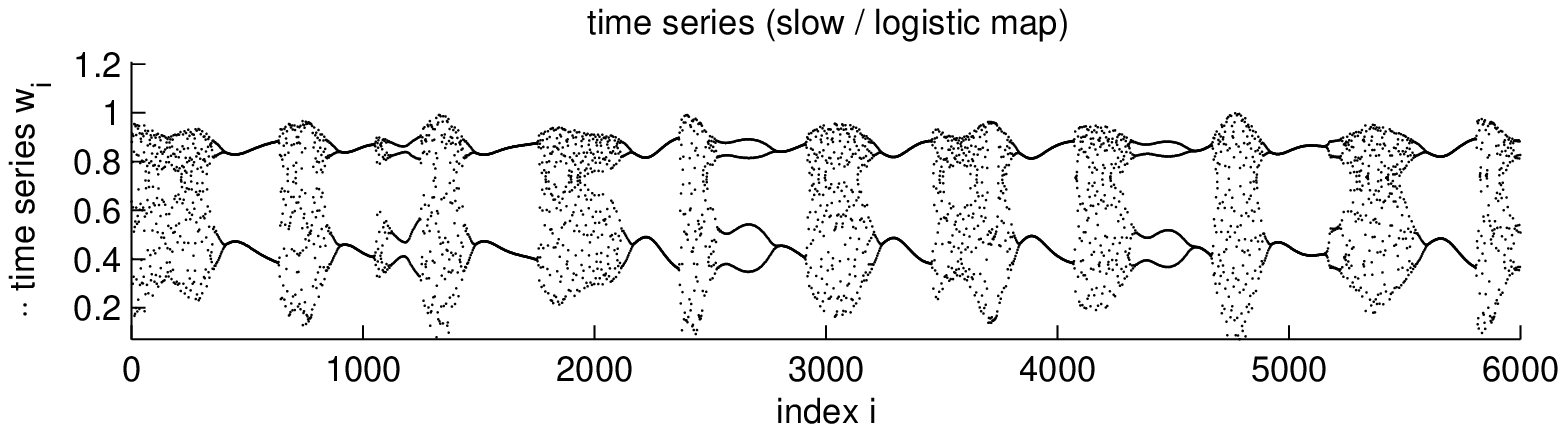}\hspace*{0.044\textwidth}}
\centerline{\includegraphics[width=0.97\textwidth]{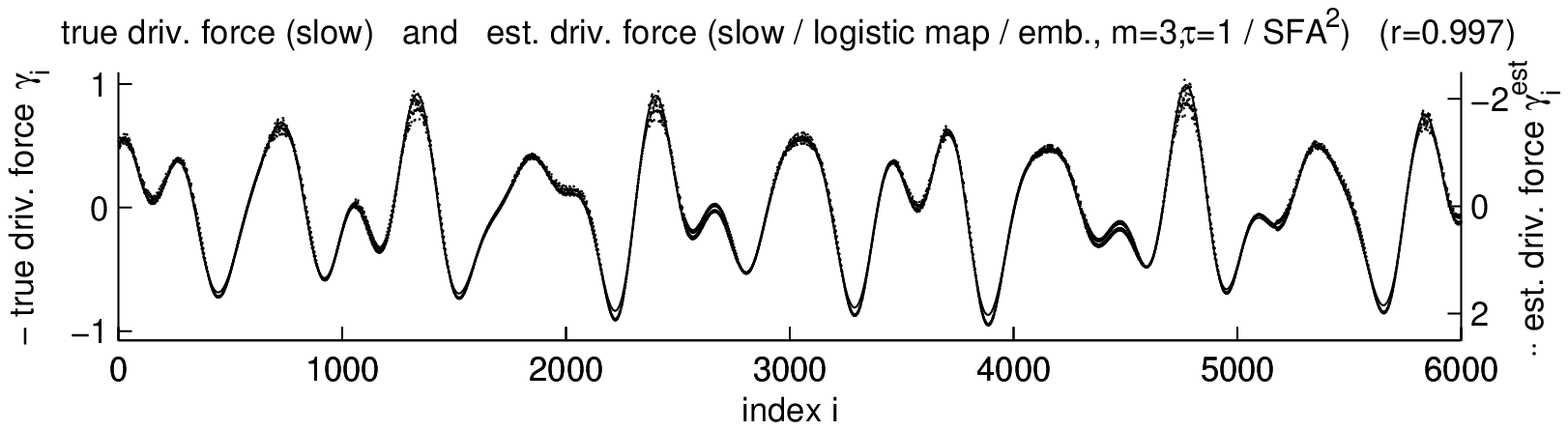}}

 \caption{\label{fig:SmoothLogistic} True slowly varying driving force
 $\p_\iu$ (bottom, solid line), time series $\uu_\iu$ (top) derived from
 the driving force with the logistic map, and estimated driving force
 $\pest_\iu$ (bottom, dots).  The correlation between true and estimated
 driving force is $r=0.997$.}

\end{figure*}

 As a second example consider a time series derived from a logistic map
\begin{equation}
  \uu_{\iu+1} = (3.6+0.4 \p_\iu) \uu_\iu (1-\uu_\iu) \,,
\end{equation}
 which maps the interval $[0,1]$ onto the interval $[0,0.9+0.1
 \p_\iu]$ and has the shape of an upside-down parabola crossing the
 abscissa at 0 and 1.  Parameter $\p$ governs the height of the parabola.
 Figure~\ref{fig:SmoothLogistic} shows the true driving force, the time
 series, and the estimated driving force with $m=3$, $\tau=1$, and second
 order polynomials for SFA.  The correlation between true and estimated
 driving force is $r=0.997$.

\section{Technical Remarks}

\paragraph{Choice of Parameters}

 SFA is basically parameter-free except for the general choice of the class
 of nonlinear basis functions.  The only other parameters to choose in this
 method are the dimension $m$ and the time delay $\tau$ of the embedding
 vectors.  The value of $\langle\dot{\os}^2\rangle$ is a fairly reliable
 indicator for a good choice of values for $m$ and $\tau$.  The smaller
 $\langle\dot{\os}^2\rangle$ the better, since there is no trivial way of
 achieving small $\langle\dot{\os}^2\rangle$-values (except if the number
 of basis functions comes close to the number of data points).  I typically
 test a certain range of $\tau$-values and successively increase the
 $m$-value until performance (measured in terms of
 $\langle\dot{\os}^2\rangle$ or $r$) is satisfactory.

\paragraph{Rarely Varying Driving Forces}

\begin{figure*}[htbp!]

\centerline{\includegraphics[width=0.926\textwidth]{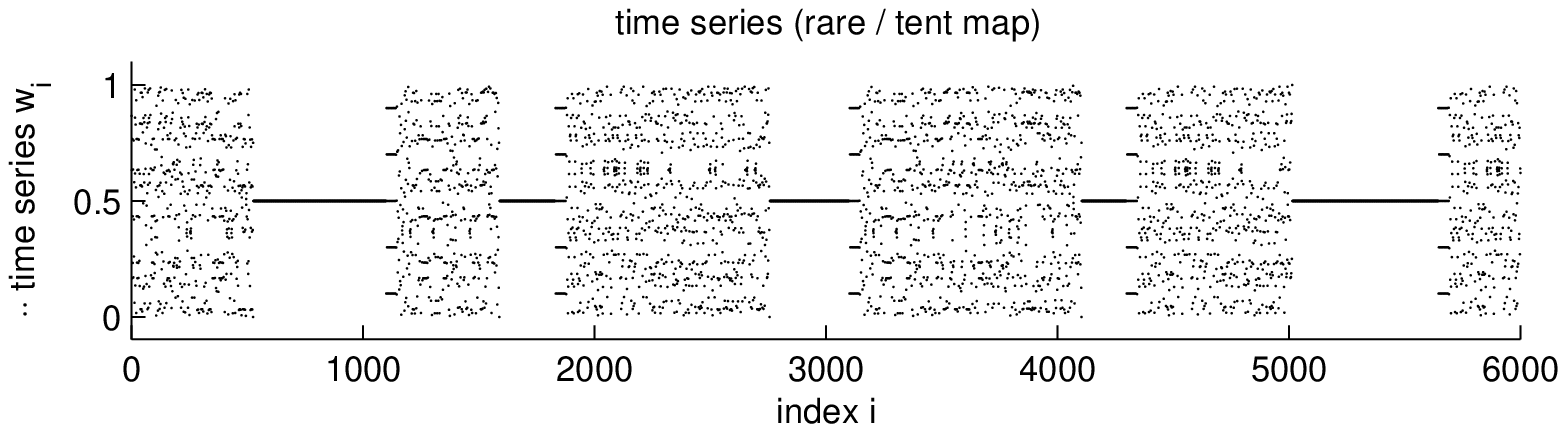}\hspace*{0.044\textwidth}}
\centerline{\includegraphics[width=0.97\textwidth]{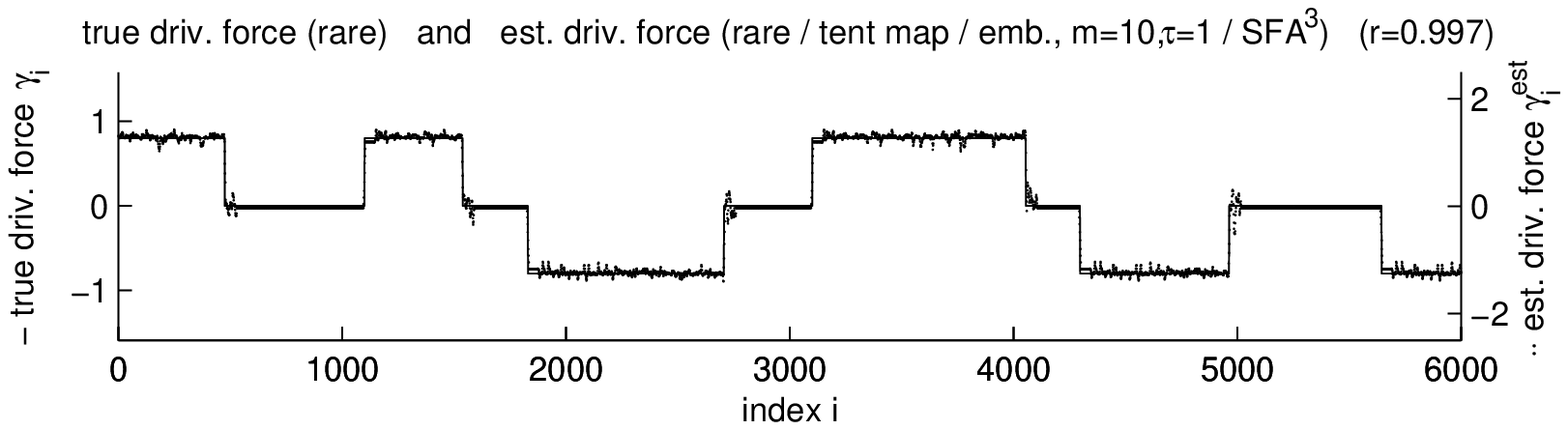}}

 \caption{\label{fig:StepsTent} True rarely varying driving force $\p_\iu$
 (bottom, solid line), time series $\uu_\iu$ (top) derived from the driving
 force with the tent map, and estimated driving force $\pest_\iu$ (bottom,
 dots).  The correlation between true and estimated driving force is $r=0.997$.}

\end{figure*}

\begin{figure*}[htbp!]

\centerline{\includegraphics[width=0.926\textwidth]{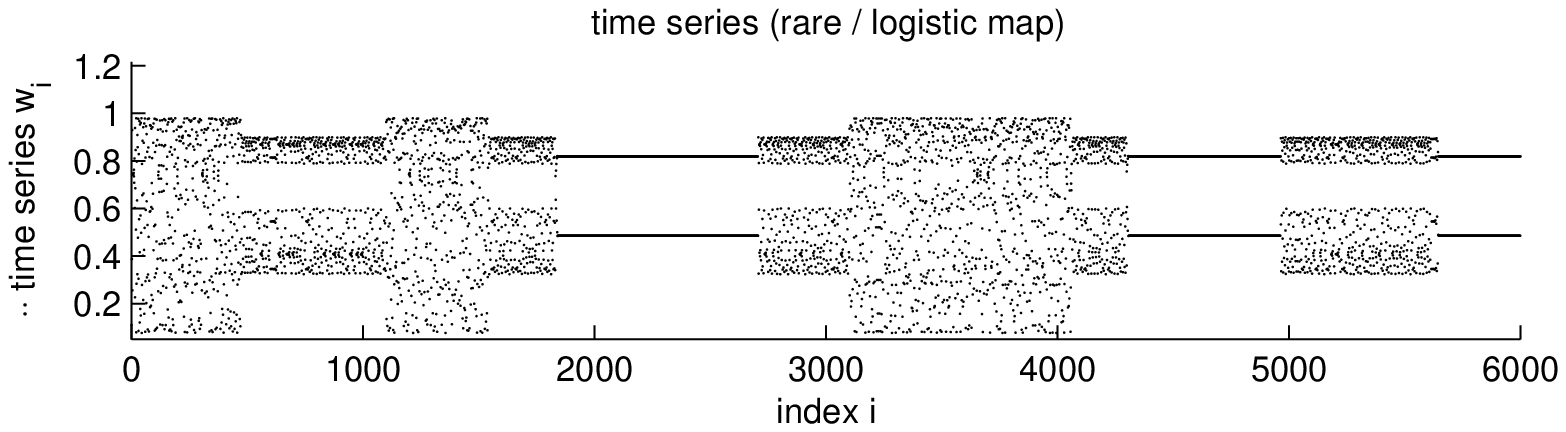}\hspace*{0.044\textwidth}}
\centerline{\includegraphics[width=0.97\textwidth]{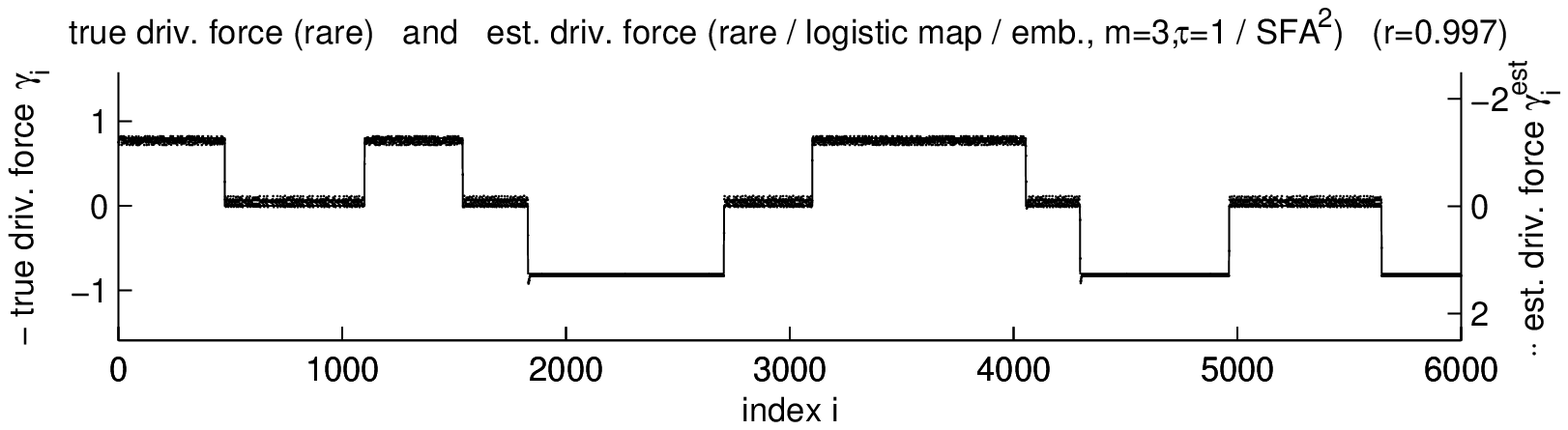}}

 \caption{\label{fig:StepsLogistic} True rarely varying driving force
 $\p_\iu$ (bottom, solid line), time series $\uu_\iu$ (top) derived from
 the driving force with the logistic map, and estimated driving force
 $\pest_\iu$ (bottom, dots).  The correlation between true and estimated
 driving force is $r=0.997$.}

\end{figure*}

 If SFA is designed to extract slowly varying features from a signal, how
 about driving forces that apparently violate this assumption heavily,
 e.g.\ if they vary abruptly and jump between different discrete values?
 The definition of slowness given in (\ref{eq:slowness}) does actually not
 depend on the smoothness of the output signal, it may equally well be a
 signal that switches abruptly between different discrete values as long as
 the jumps occur rarely enough to lead to the same variance of the time
 derivative.  Whether the order of the values can be recovered depends on
 the dynamics of the driving force.  SFA will tend to order the values such
 that jumps occur between nearby values of the estimated driving force.
 Thus, if the original values are -0.5, 0, and +0.5 and there are many
 direct jumps between -0.5 and +0.5, SFA might map the three values onto
 -1, +1, and 0, respectively, thereby changing the order of the second and
 third value.  Figures~\ref{fig:StepsTent} and~\ref{fig:StepsLogistic} show
 results for the two systems if a rarely instead of a slowly and smoothly
 varying driving force is used.

\paragraph{High-Dimensional Input Data}

 Probably the most severe limitation of SFA is the fact that the number of
 monomials (or any other basis functions) grows quickly with the
 dimensionality of the embedding vectors (curse of dimensionality).  One
 therefore might run out of computer memory before a high enough $m$-value
 is reached.  However, higher-dimensional problems could be dealt with in a
 hierarchical fashion by breaking the embedding vectors into smaller parts
 which are first analyzed separately and the results of which are then
 combined for a final analysis.  Applying this hierarchical scheme to
 65-dimensional input vectors has been demonstrated in~\cite[]{WisSej2002}.

\paragraph{Accuracy of the Estimated Driving Force}

 It might be surprising that in the examples above SFA is able to estimate
 the driving forces with such an accuracy up to a factor and a constant
 offset, even though the estimation is undetermined up to any invertible
 transformation, not only scaling and shift.  One reason for that is that
 the driving forces used are relatively slow already and could probably not
 be improved much by an invertible nonlinear transformation.  However, even
 if that were not the case one might hope that in practice for the
 lower-dimensional embedding vectors ($m=10$ instead of, e.g., $m=50$) SFA
 has to find a relatively simple input-output function $\tf(\vec{\is})$,
 which is more likely to preserve the exact shape of the driving-force
 curve than a more complicated one.  Further experiments are needed to
 verify this.

\paragraph{Noise Sensitivity}

 How sensitive is the method to noise?  One might suspect that SFA is quite
 sensitive to noise, since eigenvectors of smallest eigenvalues are used.
 However, focusing on small eigenvalues is only done after sphering and
 taking the time derivative, so that small but quickly varying noise
 components typically have large eigenvalues and are discarded by SFA.
 Graceful degradation with noise was also found in~\cite[]{WisSej2002}.
 Thus, using the eigenvectors with smallest eigenvalues in itself does not
 induce noise sensitivity.  I have done simulations with the examples
 presented here by adding Gaussian white noise to the signals before
 embedding.  Adding 10\%, 20\%, and 50\% noise to the tent map time series
 reduced the correlation between true and estimated driving force from
 $r=0.96$ to about 0.94, 0.90, and 0.71, respectively; adding 2\%, 5\%, and
 10\% noise to the logistic map time series reduced the correlation from
 $r=0.997$ to about 0.97, 0.87, and 0.70, respectively.  Thus we see, that in
 these examples the method is in fact fairly robust with respect to noise.

\paragraph{Multiple Driving Forces}

 SFA can be easily extended to the extraction of multidimensional output
 signals~\cite[]{WisSej2002}, which could be used to estimate multiple
 driving forces.  However, this might require longer time series and
 higher-dimensional embedding vectors.  If the different driving forces are
 not clearly separated by different time scales, SFA might only be able to
 estimate them up to a linear mixing transformation.  In that case
 independent component analysis (ICA) might be able to separate them, if
 they are statistically independent and not more than one is
 Gaussian~\cite[see][for an overview of ICA methods]{Hyva1999}.

\section{Conclusion}

 In this paper I have demonstrated that slow feature analysis (SFA) can be
 applied to the problem of estimating a driving force of a nonstationary
 time series.  The estimates are fairly accurate except for a scaling
 factor and a constant offset, which cannot be extracted in general.  The
 method works for slowly as well as rarely varying driving forces, although
 in the latter case the order of the different discrete values might not be
 recoverable.  The slowness principle also provides robustness to noise,
 which is typically quickly varying and therefore suppressed by SFA as far
 as possible.

 There are still a number of open questions.  The next steps of
 investigation will have to include a comparison with standard methods,
 such as the technique of recurrence plots, and an exploration of the
 conditions under which SFA can be applied with similar success as
 demonstrated here.  It is also necessary to apply SFA to real world data.
 This has been successfully done in the context of learning receptive field
 properties of the visual cortex based on natural image sequences
 \cite[]{BerkWisk2003a}, but that was not for extracting driving forces.
 In any case, since SFA works very differently from other techniques that
 have been applied to the estimation of driving forces, one can hope that
 it at least complements these other techniques in some cases.

\section{Acknowledgment}

 I am grateful to Hanspeter Herzel and Isao Tokuda for useful hints and
 Hanspeter Herzel also for critically reading the manuscript.  This work
 has been supported by the Volks\-wa\-gen\-Stif\-tung.

%\bibliographystyle{/home/wiskott/PACKAGES/TEX/BIBTEX/myapalike}
%\bibliography{/home/wiskott/BIB/bib,/home/wiskott/WWW/Bibliographies/TimeSeries}

\end{document}